# Hysteresis and compensation behaviors of mixed spin-1 and spin-2 hexagonal Ising nanowire system

**Mehmet Ertaş**[*]

*Department of Physics, Erciyes University, 38039 Kayseri, Turkey*

**Abstract**

By utilizing the framework of the effective-field theory with correlation, hysteresis and compensation behaviors of mixed spin-1 and spin-2 hexagonal Ising nanowire system is investigated in detail. The effects of Hamiltonian parameters on hysteresis behaviors are discussed, in detail. The compensation behavior of the system is also studied and according to values of Hamiltonian parameters, the Q-, S-, R- and N- type compensation behaviors is obtained in the system. Finally, the obtaining results are compared with some experimental and theoretical results and found in a qualitatively good agreement.

**1. Introduction**

In the last two decades, mixed spin Ising systems have been intensively studied. The reasons are as follows: (i) the systems have less translational symmetry than their single spin counterparts. (ii) The study of these systems can be relevant for understanding of bimetallic molecular systems based magnetic materials. (iii) Mixed spin Ising systems provide good models to investigate the ferrimagnetic materials. (iv) These systems may have a compensation temperature under certain conditions and the existence of compensation temperatures is of great technological importance such as in the thermomagnetic recording and magneto-optical readout applications [1-3]. Many combinations of the mixed spin Ising systems are possible. Some of the well-known mixed spin Ising systems are spins (1/2, 1), spins (1/2, 3/2), spins (3/2, 5/2) and spins (2, 5/2) Ising systems. The magnetic properties of these systems are studied by using different method, such as mean field theory (MFT), effective field theory (EFT), Monte Carlo (MC) simulations *etc.* (see [4-13] and references therein). Up until now, few people have ever touched upon a mixed Ising model with both spin-integer ions, namely the mixed spin-1 and spin-2 Ising system [14-20].

On the other hand, along-with the discovery of carbon nanotubes by Iijima [21], in different fields magnetic nanostructured materials have been subject of intensive interest. This is because, compared with those in bulk materials, these magnetic nanomaterials have many peculiar physical properties [22]. Another reason is that they have important potential technological applications such as medical applications, environmental remediation, permanent magnets, sensors, etc. [23-27]. Moreover, the equilibrium behavior of the nanostructures in different morphology (nanoparticle, nanowires, nanotubes ...etc.) have been a subject of extensive investigations by variety of techniques such as MFT [28, 29], EFT [30-35], and MC [36-38]. Moreover, some magnetic properties have been studied for mixed spins (1/2, 1) [39] and mixed spins (3/2, 1) [40] hexagonal Ising nanowire (HIN), ternary Ising spins (1/2, 1, 3/2) [41] and mixed spins (1/2, 3/2) [42]. We should also mention that the dynamic behavior of the nanostructures are also investigated by using EFT [43-45] and MFT [46, 47] based on Glauber dynamics and MCs [48, 49].

---

[*] Corresponding author. Tel. +90 352 2076666; Fax: +90 352 4374901.
E-mail addresses: mehmetertas@erciyes.edu.tr (Mehmet Ertaş)

Despite these studies, as far as we know, by utilizing the any method, the hysteresis and compensation behaviors of mixed spin-1 and spin-2 HIN system with core-shell structure are not studied. Therefore, in this work, we study the hysteresis and compensation behaviors of mixed spin-1 and spin-2 HIN system with core-shell structure by utilizing the EFT with correlations.

The outline of the paper is as follows. In Sec. 2, the EFT formalism is presented briefly. The detailed numerical results and discussions are given in Section 3. Finally, Section 4 is devoted to a summary and conclusion.

## 2. Formalism

The Hamiltonian of mixed spin-1 and spin-2 hexagonal Ising nanowire (HIN) system which includes nearest-neighbor interactions and crystal field is given as follows

$$H = -J_S \sum_{\langle ij \rangle} S_i S_j - J_C \sum_{\langle mn \rangle} \sigma_m \sigma_n - J_1 \sum_{\langle im \rangle} S_i \sigma_m - D \left( \sum_i S_i^2 + \sum_m \sigma_m^2 \right) - h \left( \sum_i S_i + \sum_m \sigma_m \right), \quad (1)$$

where the $J_S$ and $J_C$ are the exchange interaction parameters between two nearest neighbor magnetic atoms at the surface shell and the core, and $J_1$ is the exchange interaction between two neighbor magnetic atoms at the surface shell and the core (see Fig. 1). While the spin $\sigma$ takes $\pm 1$, 0 values, the spin S takes $\pm 2$, $\pm 1$, 0 values. <ij>, <mn> and <im> indicate summation over all pairs of nearest-neighbor sites. D and h stand for crystal field and external magnetic field, respectively. The surface exchange interaction $J_S = J_C (1+\Delta_S)$ and interfacial coupling $r = J_1/J_C$ are often defined to clarity the effects of the surface and interfacial exchange interactions on the physical properties in the nanosystem, respectively.

Within the EFT with correlations framework, one can easily find the $M_S$ and $M_C$ magnetizations, the $q_S$ and $q_C$ quadrupole moments, the $r_S$ octupolar moment and the $\upsilon_S$ hexadecapole moment as coupled equations, for the HIN system as follows:

$$\begin{Bmatrix} M_S \\ q_S \\ r_S \\ \upsilon_S \end{Bmatrix} = [a_0 + a_1 M_S + a_2 q_S + a_3 r_S + a_4 \upsilon_S]^4 [1 + M_C \sinh(J_1 \nabla) + q_C (\cosh(J_1 \nabla) - 1)] \begin{bmatrix} F_m(x)|_{x=0} \\ F_q(x)|_{x=0} \\ F_r(x)|_{x=0} \\ F_\upsilon(x)|_{x=0} \end{bmatrix}, \quad (2a)$$

$$\begin{Bmatrix} M_C \\ q_C \end{Bmatrix} = [1 + M_C \sinh(J_C \nabla) + q_C (\cosh(J_C \nabla) - 1)]^2 [b_0 + b_1 M_S + b_2 q_S + b_3 r_S + b_4 \upsilon_S]^6 \begin{bmatrix} G_m(x)|_{x=0} \\ G_q(x)|_{x=0} \end{bmatrix}. \quad (2b)$$

Here, we give $a_i$ and $b_i$ coefficients in the Appendix and define the F(x) and G(x) functions as follows:

$$F_m(x) = \frac{1}{2} \frac{4 \sinh[2\beta(x+h)] + 2 \sinh[\beta(x+h)] \exp(-3\beta D)}{\cosh[2\beta(x+h)] + \cosh[\beta(x+h)] \exp(-3\beta D) + \exp(-4\beta D)}, \quad (3a)$$

$$F_q(x) = \frac{1}{2} \frac{8\sinh[2\beta(x+h)] + 2\sinh[\beta(x+h)]\exp(-3\beta D)}{\cosh[2\beta(x+h)] + \cosh[\beta(x+h)]\exp(-3\beta D) + \exp(-4\beta D)}, \quad (3b)$$

$$F_r(x) = \frac{1}{2} \frac{16\sinh[2\beta(x+h)] + 2\sinh[\beta(x+h)]\exp(-3\beta D)}{\cosh[2\beta(x+h)] + \cosh[\beta(x+h)]\exp(-3\beta D) + \exp(-4\beta D)}, \quad (3c)$$

$$F_\upsilon(x) = \frac{1}{2} \frac{32\sinh[2\beta(x+h)] + 2\sinh[\beta(x+h)]\exp(-3\beta D)}{\cosh[2\beta(x+h)] + \cosh[\beta(x+h)]\exp(-3\beta D) + \exp(-4\beta D)}, \quad (3d)$$

and

$$G_m(x) = \frac{2\sinh[\beta(x+h)]}{2\cosh[\beta(x+h)] + \exp(-\beta D)}, \quad (4a)$$

$$G_q(x) = \frac{2\cosh[\beta(x+h)]}{2\cosh[\beta(x+h)] + \exp(-\beta D)}, \quad (4b)$$

where $\beta = 1/kT$, k is the Boltzmann constant and k = 1.0 throughout the paper, T is the absolute temperature. By utilizing the definitions of the magnetizations Eqs. (2a) and (2b), we can define the $M_T$ total magnetization of each site from Fig. 1 as $M_T = 1/7(6M_S + M_C)$. It is worthwhile mentioning that the thermal behaviors of $q_S$, $q_C$, $r_S$ and $r_C$ were not studied. Because, the Hamiltonian of the HIN system did not include the K biquadratic exchange interaction parameter. However, we need Eqs. (2a) and (2b) to determine the behaviors of $M_S$ and $M_C$. Solving these equations, the hysteresis and compensation behaviors of mixed spin-1 and spin-2 HIN system with core-shell structure can be obtained and in the next Section, the calculations will be performed in detail.

### 3. Numerical results and discussions

In this section, we present the numerical calculations for the hysteresis and compensation behaviors of mixed spin-1 and spin-2 HIN system with core-shell structure in the following figures, namely Figs. 2-6.

#### *3.1. Hysteresis behaviors of mixed spin-1 and spin-2 HIN system*

In this subsection, the effects of the temperature, crystal field and interlayer coupling are investigated on the hysteresis behaviors of mixed spin-1 and spin-2 HIN system.

#### *3.1.1. The effects of the temperature*

The temperature dependence of hysteresis loops of mixed spin-1 and spin-2 HIN system will be discussed in Subsection. Figs. 2(a)-(e) illustrate the total magnetization $M_T$, the surface shell magnetization $M_S$, and the core magnetization $M_C$ versus external field for r=1.0, D=1.0, $\Delta_S$= 0.0 and the five different values of T, namely T = 1.0, 5.0, 7.0, 9.0, 10.0. The HIN system displays a very hard magnet with a wide hysteresis loop at low temperatures. The system exhibits a single hysteresis loop as seen in Figs. 2(a)-(d) and when approaching the critical temperature value, the single hysteresis loop turns out thinner and narrower. Furthermore, as seen Fig. 2(e), if the temperature increases stronger, the single hysteresis loop disappears. These results are in a good agreement with the theoretical results [50, 51] and the experimental results [52, 53], namely, $La_{2/3}Sr_{1/3}MnO_3$ nanoparticle assembled nanotubes [52], and FeGa/CoFeB, FeGa/Py, CoFeB/Cu, Py/Cu, CoNiP/Cu multilayered nanowires [53].

### *3.1.2. The effects of the crystal field*

The crystal field dependence of hysteresis loops of mixed spin-1 and spin-2 HIN system will be discussed in Subsection. Figs. 3(a) and 3(b) are calculated for r = 1.0, $\Delta_S$= 0.0 fixed values, two different temperature which is low and high values and five different crystal field values, namely T = 0.5 and D= 0.0, -1.0, -1.5, -2.0, -4.0 as seen Fig. 3(a) and, T = 2.0 and D= 0.0, -1.0, -1.5, -2.0, -3.0 as seen Fig. 3(b). Fig. 3(a) illustrates the crystal field dependence of hysteresis loops for the low temperature value. In this figure, we can see that with decreasing crystal field, the shape of hysteresis loops becomes smaller and the HIN system displays a single hysteresis loop for D= 0.0, -1.0, -1.5, -2.0 values. Moreover, at D= -4.0 value, the single hysteresis loop turns into a double hysteresis loops. The hysteresis loops have been observed $Fe_3O_4/ZrO_2/Fe_3O_4$ multilayer nanotube [54], experimentally. Theoretically, similar result has been observed spin-3/2 cylindrical Ising nanotube [55] and spin-3/2 HIN system [34]. Fig. 3(b) displays the crystal field dependence of hysteresis loops for the high temperature value. Fig. 3(b) is similar to Fig. 3(a), except for the following differences: (1) the system in the Fig. 3(b) does not exist a double hysteresis loops. (2) If the crystal field decreases stronger, the single hysteresis loop disappears in Fig. 3(b).

### *3.1.3. The effect of ferromagnetic and antiferromagnetic interlayer coupling*

Fig. 4 and Fig. 5 are calculated for examine the effect of ferromagnetic and antiferromagnetic interlayer coupling, respectively. Fig. 4 is calculated for T = 0.50, D = -1.0, $\Delta_S$ = 0.0 and six different ferromagnetic interlayer coupling values, namely r = 0.01, 0.1, 0.25, 0.4, 0.7, 1.0. For the low values of the ferromagnetic interlayer coupling, the system occurs a double hysteresis loops as seen Fig. 4(a)-(c). Moreover, for the high values of the ferromagnetic interlayer coupling, a double hysteresis loops turn out one hysteresis loop as seen Figs. 4(d)-(f) and the single hysteresis loop turns out thicker and wider with increasing ferromagnetic interlayer coupling. Fig. 5 is performed for T = 0.50, D = -1.0, $\Delta_S$ = 0.0 and six different ferromagnetic interlayer coupling values, namely r = -0.01, -0.1, -0.25, -0.4, -0.7, -1.0. For the high values of the antiferromagnetic interlayer coupling, the system occurs a double hysteresis loops as seen Fig. 5(a)-(c). Moreover, for the low values of the antiferromagnetic interlayer coupling, a double hysteresis loops turn out one hysteresis loop as seen Figs. 5(d)-(f) and the single hysteresis loop turns out thicker and wider with decreasing antiferromagnetic interlayer coupling.

### *3.2. Compensation behaviors of mixed spin-1 and spin-2 HIN system*

As known, the existence of compensation temperatures is of great technological importance such as in the thermomagnetic recording and magneto-optical readout applications [1-3]. To this end, we also investigated the temperature variation of the total magnetization, the core magnetization and the surface shell magnetization for various values of Hamiltonian to obtain the compensation temperature and determine compensation types as seen in Fig. 6. Fig. 6(a) is calculated for r = -0.5, $\Delta_S$ = 0.0, D = 0.0 and the HIN system displays the Q type of compensation behaviors. The HIN system exhibits the S-type of compensation behaviors for r = -0.5, $\Delta_S$ =-0.99, D = 0.0 values as seen Fig. 6(b). The R-type behavior is observed in Fig. 6(c) for r = 1.0, $\Delta_S$ = -0.5, D = 0.0. Fig. 6(d) illustrates the P-type behaviors for r = -0.3,

$\Delta_S = -0.99$, $D = 0.0$. The Q-, R-, and N- types of compensations behaviors classified in the Néel theory [56] and S-type was obtained by Strečka [57].

## 4. Summary and conclusion

By utilizing the EFT, we studied the hysteresis and compensation behaviors of mixed spin-1 and spin-2 HIN and investigated the effects of Hamiltonian parameters on hysteresis in detail. The compensation behavior of the HIN system is also examined and according to values of Hamiltonian parameters, the Q-, S-, R- and N- type compensation behaviors is obtained in the HIN system. Finally, the obtaining results are compared with some experimental and theoretical results and found in a qualitatively good agreement.

**Appendix**

$a_0 = 1$,

$a_1 = \frac{1}{6}\left[8\sinh(J_S\nabla) - \sinh(2J_S\nabla)\right]$,

$a_2 = \frac{1}{12}\left[16\cosh(J_S\nabla) - \cosh(2J_S\nabla) - 15\right]$,

$a_3 = \frac{1}{6}\left[\sinh(2J_S\nabla) - 2\sinh(J_S\nabla)\right]$,

$a_4 = \frac{1}{12}\left[\cosh(2J_S\nabla) - 4\cosh(J_S\nabla) + 3\right]$,

$b_0 = 1$,

$b_1 = \frac{1}{6}\left[8\sinh(J_I\nabla) - \sinh(2J_I\nabla)\right]$,

$b_2 = \frac{1}{12}\left[16\cosh(J_I\nabla) - \cosh(2J_I\nabla) - 15\right]$,

$b_3 = \frac{1}{6}\left[\sinh(2J_I\nabla) - 2\sinh(J_I\nabla)\right]$,

$a_4 = \frac{1}{12}\left[\cosh(2J_I\nabla) - 4\cosh(J_I\nabla) + 3\right]$,

**Figure Captions**

**Fig. 1.** (Color online) Schematic representation of hexagonal Ising nanowire. The blue and red spheres indicate magnetic atoms at the surface shell and core, respectively.

**Fig. 2.** (Color online) Thermal variations of the magnetizations with the fixed values of r =1.0, D=1.0, $\Delta_S$= 0.0 and the various values of T; **(a)** T = 1.0, **(b)** T= 5.0, **(c)** T = 7.0, **(d)** T = 9.0, **(e)** T = 10.0.

**Fig. 3.** (Color online) Hysteresis behaviors of the mixed spin-1 and spin-2 HIN system. **(a)** r = 1.0, $\Delta_S = 0.0$, T = 0.5 and for various values of D, namely, D = 0.0, -1.0, -1.5, -2.0, -4.0, **(b)** r = 1.0, $\Delta S = 0.0$, T = 2.0 and for various values of D, namely, D = 0.0, -1.0, -1.5, -2.0, -3.0,

**Fig. 4.** Same as Fig. 3, but for T = 0.50, D = -1.0, $\Delta_S = 0.0$, and

**(a)** r = 0.01; **(b)** r = 0.1; **(c)** r = 0.25; **(d)** r = 0.4; **(e)** r = 0.7; **(f)** 1.0.

**Fig. 5.** Same as Fig. 3, but for T = 0.50, D = -1.0, $\Delta_S = 0.0$, and

**(a)** r = -0.01; **(b)** r = -0.1; **(c)** r = -0.25; **(d)** r = -0.4; **(e)** r = -0.7; **(f)** -1.0.

**Fig. 6.** (Color online) The total magnetization, core magnetization and shell magnetization as a function of the temperature for different values of interaction parameters. The system exhibits the Q-, S-, R-, and N- types of compensation behaviors.

**(a)** r = -0.5, $\Delta_S = 0.0$, D = 0.0; **(b)** r = -0.5, $\Delta_S = -0.99$, D = 0.0; **(c)** r = 1.0, $\Delta_S = -0.5$, D = 0.0; **(d)** r = -0.3, $\Delta S = -0.99$, D = 0.0.

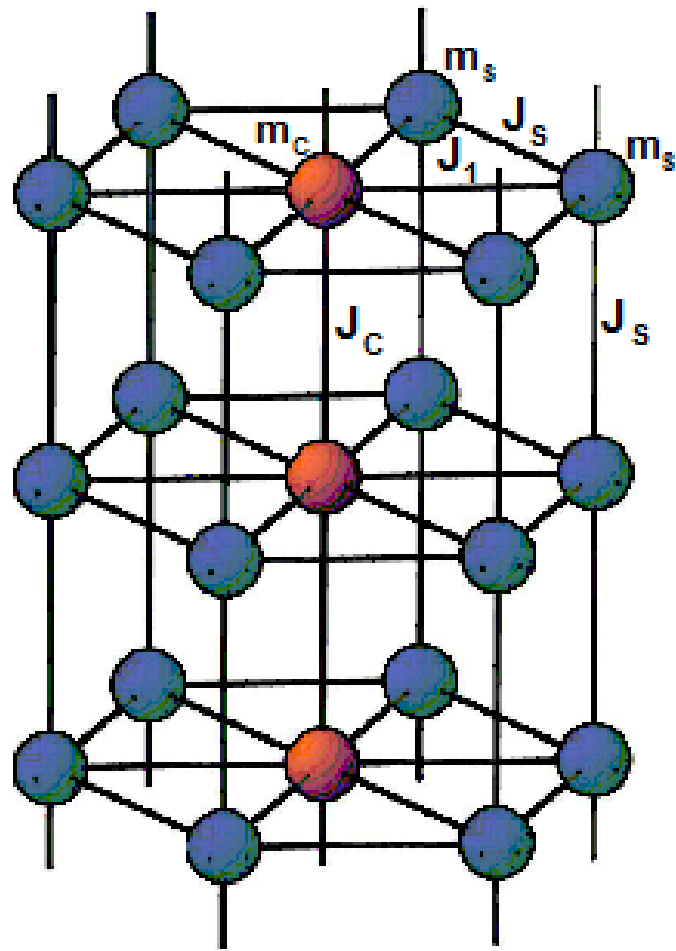

**Fig. 1**

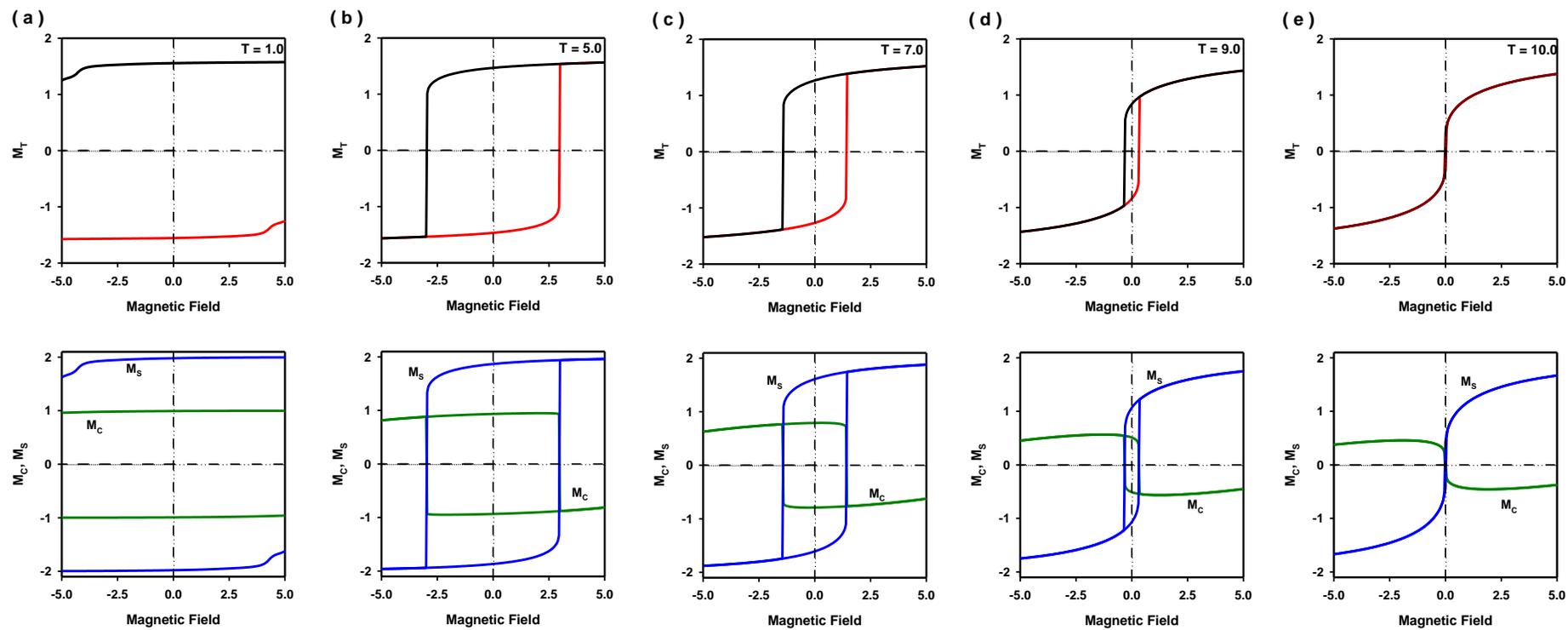

**Fig. 2**

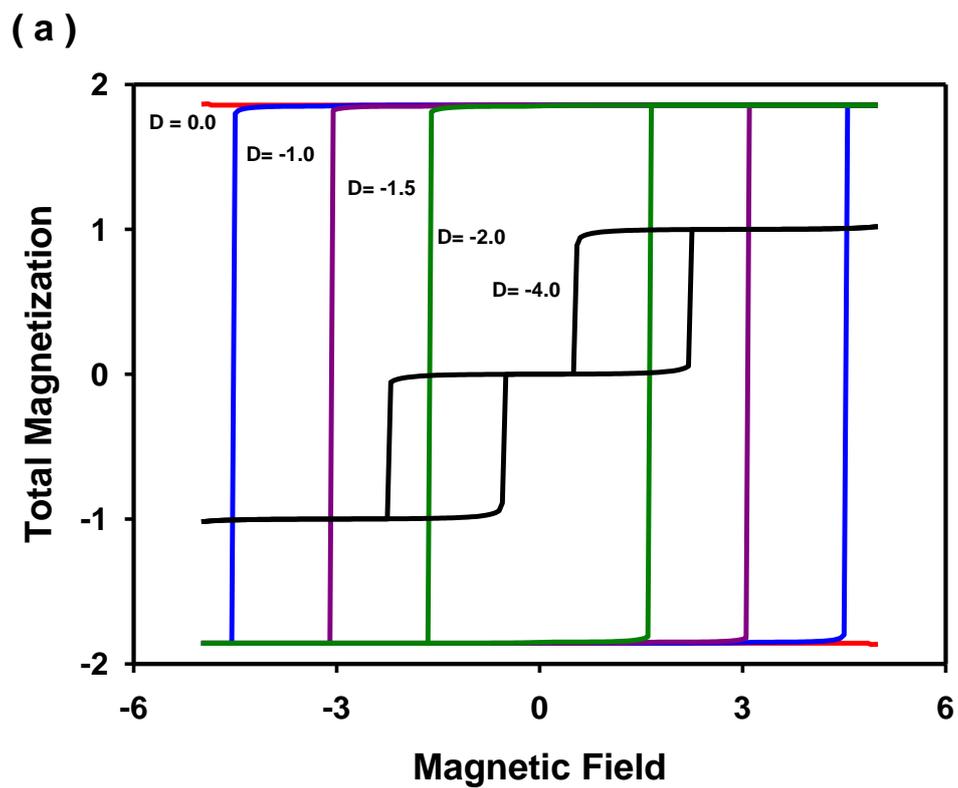

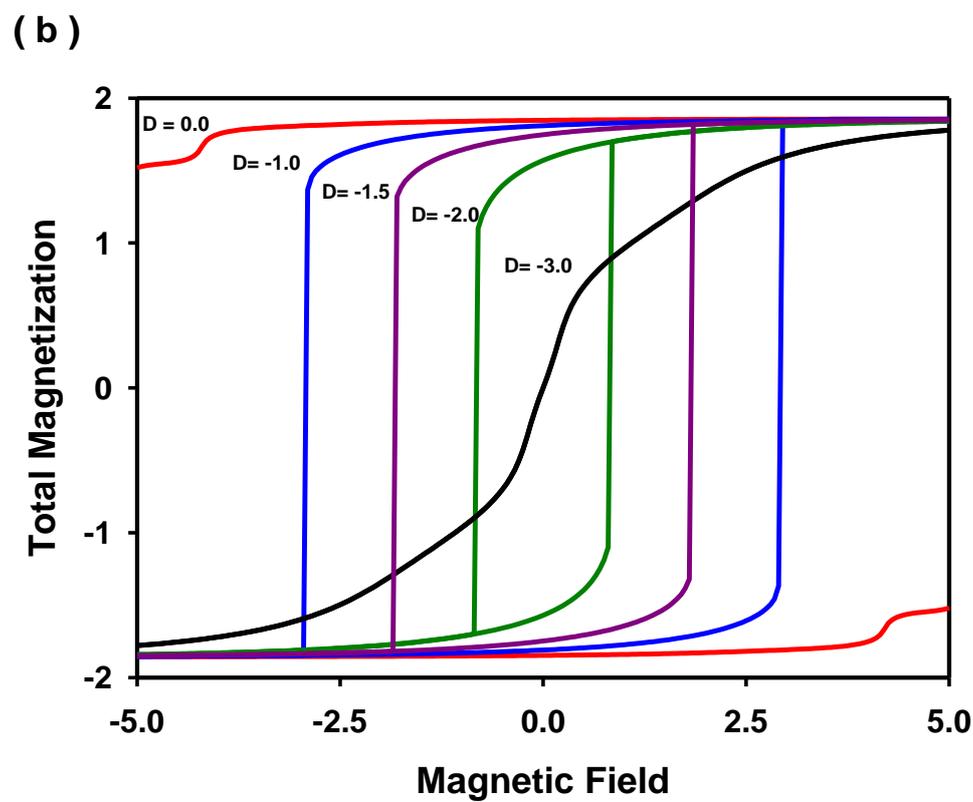

**Fig. 3**

**Figure 4**

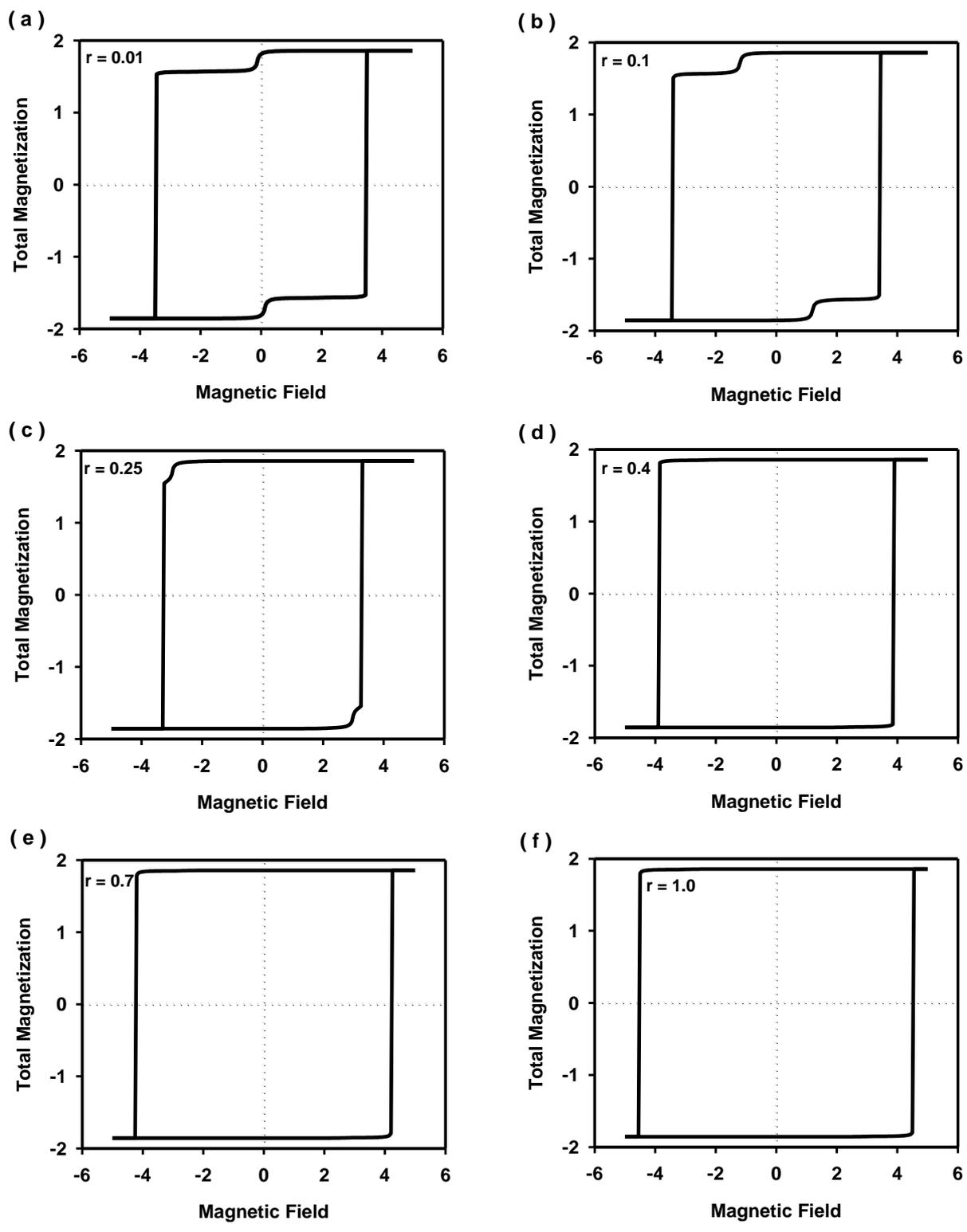

**Fig. 4**

**Figure 5**

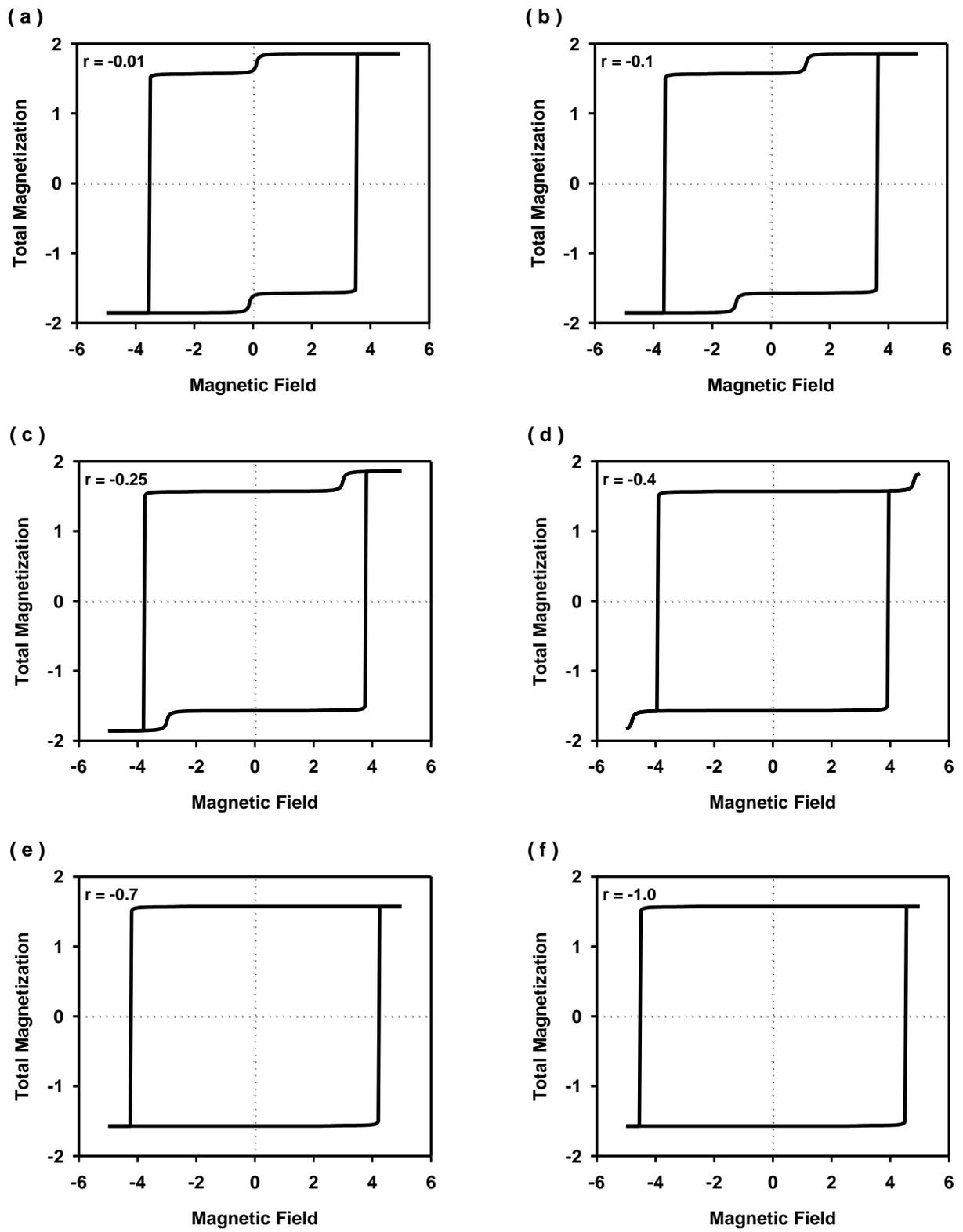

**Fig. 5**

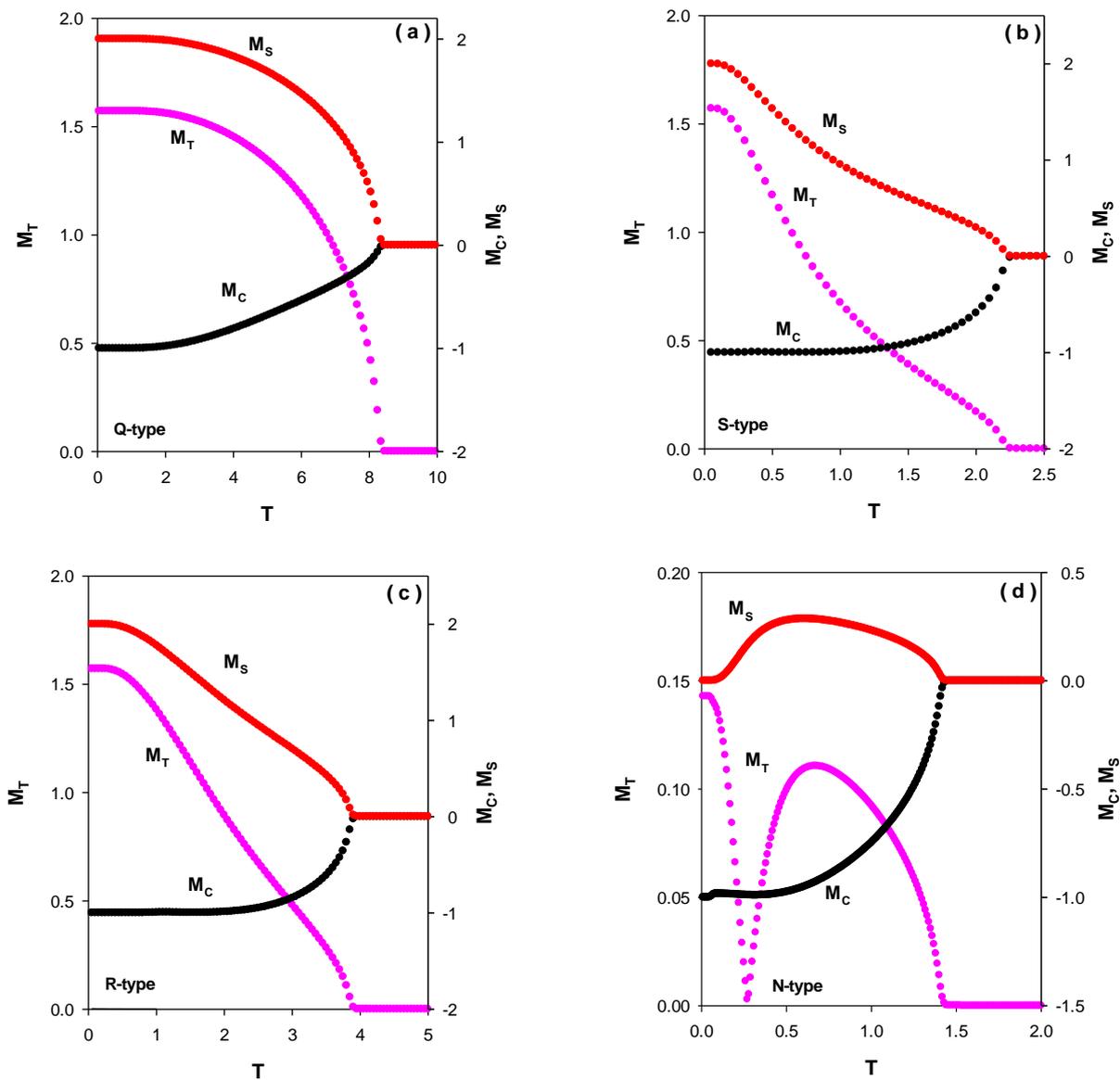

**Fig. 6**

*Highlights (for review)

# Research Highlights

- ➢ Magnetic properties of a mixed hexagonal Ising nanowire are studied.
- ➢ Hysteresis behaviors are investigated, in detail.
- ➢ Compensation behaviors are studied, in detail.
- ➢ We examined the effects of the Hamiltonian parameters on the system.